%% file: gcd-mult22.tex
\documentclass[aps,pra,twocolumn,showpacs,balancelastpage,superscriptaddress,groupedaddress]{revtex4}  % for review and submission

\usepackage{color}
\usepackage[T1]{fontenc}
\usepackage[latin1]{inputenc}
\usepackage[english]{babel}
\usepackage[full]{complexity}
\usepackage{hyperref}

\usepackage{graphicx,subfigure}  % needed for figures
\usepackage{dcolumn}   % needed for some tables
\usepackage{bm}        % for math
\usepackage{amssymb}   % for math
\input{Qcircuit.tex}

\renewcommand{\Qcircuit}[1][0em]{\xymatrix @*=<#1>}
\newcommand{\CM}{${\mathcal CM}$}
\newcommand{\UM}{${\mathcal UM}$}
\newcommand{\hide}[1]{}
\begin{document}

\title{\large Faster Quantum Number Factoring via Circuit Synthesis}
\author{Igor L. Markov}%
\email[Address correspondence to: ]{imarkov@eecs.umich.edu}
\affiliation{Department of EECS, University of Michigan, Ann Arbor, MI 48109-2121}

\author{Mehdi Saeedi}%
\affiliation{Department of Electrical Engineering, University of Southern California, Los Angeles, CA 90089-2562}

%\date{}
\begin{abstract}
A major obstacle to implementing Shor's quantum number-factoring algorithm
is the large size of modular-exponentiation circuits. We reduce this bottleneck by customizing reversible circuits for modular multiplication to individual runs of Shor's algorithm. Our circuit-synthesis procedure {exploits spectral properties of  multiplication operators and constructs optimized} circuits from the traces
of the execution of an appropriate GCD algorithm. Empirically, gate counts are reduced
by 4-5 times, and circuit latency is reduced by larger factors.
\end{abstract}
%Use showkeys class option if keyword

\maketitle

\section{Introduction}
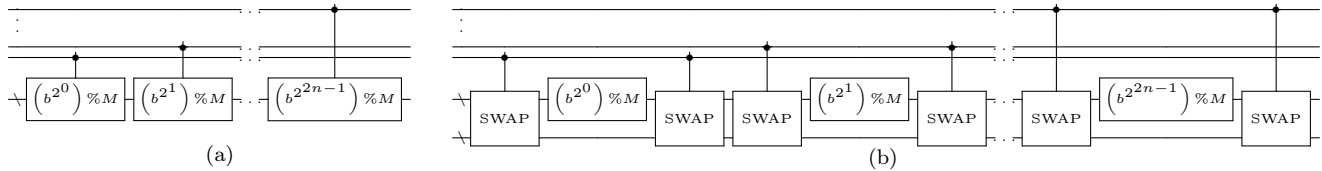
\begin{figure*}[t]
        \tiny
        \subfigure[\label{fig:shor1}]{
        \input{Shor}
        }\subfigure[\label{fig:shor2}]{
        \input{mux}
        }
        \vspace{-4mm}
        \caption{(a) \label{fig:shor}
         Modular exponentiation using conditional modular multiplications \cite{Beckman1996}. (b) Conditional multiplications implemented using unmodified unconditional modular multiplication blocks and conditional swaps with a zero register \cite{modmultqic}.}
                 \vspace{-2mm}
\end{figure*}

\vspace{-2mm}
Shor's number factoring remains the most striking algorithm for quantum computation as it quickly solves an important task~\cite{NielsenC2001} for which no conventional fast algorithms were found in 2,300 years~\footnote{Euclid studied number factorization circa 300 B.C.E. as a way to compute GCD, which is required to add fractions. Failing to find a fast algorithm, he developed what is now known as Euclid's GCD algorithm.}. Today, a scalable implementation of Shor's technique would have dire implications to Internet commerce. Laboratory demonstrations factored $15=3\cdot5$ circa 2000 \cite{f15}, but further progress was slow \cite{Lanyon2007,LuBYP2007,PolitiMB09} as
 factoring sizable semiprimes requires very large circuits. The bottleneck of Shor's number factoring is in {\em modular exponentiation} --- a reversible circuit computing ($b^z \mod M$) for known coprime integers $b$ and $M$. This computation is performed as a sequence of conditional
modular multiplications (\CM) \cite{Beckman1996} by pre-computed powers of a randomly selected base value ($b$), controlled by the bits of $z$ (Figure \ref{fig:shor}).
In most cases, $b=2$ or $b=3$ suffice \cite{modmultqic}. Such \CM~blocks are assembled
from unmodified unconditional modular multiplication blocks (\UM) using pre- and
post-processing: since multiplication always preserves the integer 0, a \UM~block can be
"turned off" by conditionally swapping a 0 with its inputs and then restoring the inputs
by an identical swap. Conditional swaps can be simplified, and further circuit optimizations focus on \UM~blocks. These steps are reviewed in detail in \cite{modmultqic}.

\vspace{-2mm}
 \section{Prior work}
 \label{sec:prior}
\vspace{-2mm}
 \UM~blocks are assembled from modular additions and multiplications by two, scheduled according to the binary expansion of the constant multiplicand and its modular inverse~\cite{Beckman1996} (see a contemporary summary in \cite{modmultqic}). In one popular approach, the input value $x$ is copied into a zero-initialized register to obtain $(x,x)$. To compute $13x$, follow the binary expansion 13=$b$1101: $(x,x)$-$(2x,x)$-$(3x,x)$-$(6x,x)$-$(12x,x)$-$(13x,x)$. Now the second register must be restored to 0 for the next \UM~block to use it. However, this requires dividing $13x$ by 13, i.e., multiplying by the {\em modular inverse} of 13. For $M=101113=569\cdot1777$, the inverse of $C=13$ is 77778, (10010111111010010$_2$), requiring a large circuit. In \cite{modmultqic}, we constructed alternative circuits without computing modular inverses. To accomplish this, we introduced circuit blocks for modular multiplication and division by two that restore their ancillae to 0. We then estimated costs of circuit blocks for modular addition, subtraction, multiplication and division by two, and several others \cite[Table 2]{modmultqic}. Using these blocks, we found optimal \UM~circuits for each $C,M$ up to 15 bits. The same procedure can be used for different cost estimates, but optimal search does not scale well beyond 15 bits. Numerical results demonstrated that traditional circuits based on binary expansion are far from optimal, thus asking for scalable constructions beyond 15 bits.

  Researchers optimizing circuits for Shor's algorithm \cite{Knill95,McAnally01} adapted these circuits to use only nearest-neighbor quantum couplings \cite{Fowler2004} and restructured them to leverage parallel processing \cite{VanMeter2005}.
  Applying multiple quantum couplings in parallel allows one to finish computation faster. The smaller required lifespan of individual qubits additionally reduces the susceptibility of qubits to decoherence and decreases the overall need for quantum error-correction.
  The runtime (latency) of parallel quantum computation is estimated by the {\em depth} of its quantum circuit, i.e., the maximum number of gates on any input-output path. Depth reductions in the literature sharply increase the required number of qubits, e.g., by 50 times or more, making them impractical for modern experimental environments where controlling 50-100 qubits remains a challenge. Vice versa, prior circuits with 1-2 fewer qubits use more gates \cite{Takahashi2010}. Rosenbaum has shown \cite{Rosenbaum12} how to adapt unrestricted circuits to nearest-neighbor architecture using teleportation, while asymptotically preserving their depth. For Shor's algorithm, the range of practical interest is currently between several hundred and a thousand logical qubits, where FFT-based multiplication needs more gates than simpler techniques.

  Our circuits moderately increase qubit counts to significantly decrease gate counts and circuit depth. Built from standard components, they are readily adapted to nearest-neighbor quantum architectures by optimizing these components to each particular architecture~\cite{Fowler2004}.

\section{New Circuits}
\vspace{-2mm}
 We propose two-register \UM~circuits to compute $(Cx \mod M)$ for coprime $C$ and $M$, GCD$(C,M) = 1$. These circuits transform $(x,0)$ into $(x,x)$ using
 parallel CNOT gates and then compute $(Cx\mod M,0)$. Clearing ancillae in the second register allows the next  circuit module to use them again. As reversible building blocks, we use {\em modular addition and subtraction between the two registers} $(a,b)\mapsto(a\pm b \mod M,b)$ and $(a,b)\mapsto(a, a\pm b \mod M)$, as well as circuits from \cite{modmultqic} for modular multiplication by two that clear their ancillae $a \mapsto 2a\mod M$.

 Our key insight is to use the {\em coprimality} of $C$ and $M$ (guaranteed in Shor's algorithm) to read off a circuit from the execution trace of an appropriate GCD algorithm. Recall that the Euclidean GCD algorithm for $(A,B)$ proceeds by replacing the larger number $A$ with $(A \mod B)$ until the result evaluates to 0. For $C=13$ and $M=21$, this produces the Fibonacci sequence
 $(21,13)$-$(8,13)$-$(8,5)$-$(3,5)$-$(3,2)$-$(1,2)$-$(1,0)$. For convenience, one may consider the last configuration to be $(1,1)$, so that each step performs a subtraction --- a simpler operation than modular reduction. As a result, we obtain GCD$(C,M)=1$. Reversing the order of operations, interpreting each number as a multiple of $x$ starting with $(1x,1x)$, and mapping each step into a mod-21 addition, we obtain
 $(x, x)$-$(2x, x)$-$(2x, 3x)$-$(5x,3x)$-$(5x, 8x)$-$(13x, 8x)$-$(13x, 21x)=(13x,0)$.
 Since $21x \mod 21=0$, the second register is restored to 0. This
\UM~circuit bypasses Bennett's construction based on modular inverses (Section \ref{sec:prior})
and is smaller than prior art \cite{NielsenC2001,Beckman1996}.

Unfortunately, some modular reductions in the Euclidean GCD algorithm may require a large number of gates. Consider $(11x \mod 21)$ and its Euclidean GCD trace $(21,11)$-$(10,11)$-$(10,1)$-$(1,1)$.
Implementing the last $\mod$ operation by nine subtractions produces a sequence of nine mod-21 additions $(x,x)$-$(2x,x)$-$(3x,x)$-$\ldots$-$(10x,x)$.
To improve efficiency, we replace the Euclidean GCD algorithm by a binary GCD algorithm that avoids the $\mod$ operation and uses a shortcut for the case of odd GCD. Given a pair of odd numbers, the larger one is replaced by their difference, which must be even. Any even number is divided by two, which can be implemented by a controlled bit-shift (as shown in \cite{modmultqic}).\\
  For even $A$ and $B$, $~~~~~~(A,B)=(A/2, B/2)$ \\
  For even $A$ and odd $B$, $(A,B) = (A/2, B)$ \\
  For odd  $A$ and even $B$, $(A,B) = (A, B/2)$ \\
  For odd  $A$ and $B$, $~~~~$ if $A<B$, $(A,B)=(A,B-A)$ \\
   ~~~~~~~~~~~~~~~~~~~~~~~~~~ else $~~~~~~~~(A,B)=(A-B,B)$ \\
One stops when $A$=$B$=GCD=1 (assuming coprime inputs). The sequence of operations performed for our example $(21,11)$-$(10,11)$-$(5,11)$-$(5,6)$-$(5,3)$-$(2,3)$-$(1,3)$-$(1,2)$-$(1,1)$
can be improved by $(2,3)$-$(2,1)$-$(1,1)$. To obtain a circuit, such sequences are reversed and interpreted as modular multiplications by two and modular additions, with the initial state $(1x,1x)$.
Further improvements are obtained by allowing both subtractions and additions, e.g., $15x=16x-x$ versus $15x=8x+4x+2x+1x$ (here $16x$, $8x$, etc are computed by doubling).

In a more involved example $(7x \mod 1017)$, the addition leading to $(7,1024)$ is a better first step than the subtraction leading to $(7,1010)$, because $(7,1024)$ enables eight successive divisions by two which reduce the values down to $(7,4)$ faster than subtractions would. Then, subtractions become the best operators: $(3,4)$-$(3,1)$-$(2,1)$-$(1,1)$. This optimization relies on a three-step lookahead. To select each next operator, we consider all possible irredundant three-step sequences of operators (modular addition, subtraction and division by two), find their final states, and score the remaining circuit according to the trace of the binary GCD algorithm (without lookahead). The cost of each operator/step can be specific to the quantum machine. Taking the best three-step sequence, we commit to its first operator. The remaining two steps are ignored, and the next operator is chosen by a separate round of lookahead. For $(11x \mod 21)$, we obtain $(21,11)$-$(10,11)$-$(5,11)$-$(5,6)$-$(5,1)$-$(4,1)$-$(2,1)$-$(1,1)$.

\begin{table*}
\caption{\label{tab:avg_max}  Circuits produced by our technique and prior art, compared by Toffoli gate counts. Circuit sizes for $n<16$ are averaged over all $M$-coprime $C$ values. Results for $n=16$ include all coprime $C$ values for the given $M$. For 24-, 32-, 48-, and 64-bit $M$ values, results are averaged over the first 5000 coprime $C$ values. For larger $n$ values in mod-mult circuits, only $C=-1/17 \mod M$ (e.g., 47679095568306588235294117647058823529411764705882352941176470588235294117647 for $n=256$) are shown. For modular exponentiation, results include all $C$ values appearing in \UM~blocks for $b=2$. All results reported are circuit sizes (Toffoli gate counts), except for values in the \emph{depth} column. For `Avg ratio' in mod-mult, we used Ours/\cite{modmultqic} and \cite{Beckman1996}/Ours.}
\scriptsize
 \begin{tabular}{|c|l||cc|cc|cc|cc||cc|c|c|}
  \hline
   & & \multicolumn{8}{c||}{Modular Multiplication (circuit size)} & \multicolumn{4}{c|}{Modular Exponentiation} \\
 Bits& \# of semiprimes & \multicolumn{2}{c|}{Optimal \cite{modmultqic}} & \multicolumn{2}{c|}{Ours} & \multicolumn{2}{c|}{Beckman \cite{Beckman1996}} & \multicolumn{2}{c||}{Avg ratio}  &  \multicolumn{2}{c|}{Ours} & Beckman & \multicolumn{1}{c|}{Avg ratio} \\
$n$ & [smallest, largest] & max & avg & max & avg  & max & avg  & /\cite{modmultqic} & \cite{Beckman1996}/ & \#Gates & Depth &  \cite{Beckman1996} & \cite{Beckman1996}/\rm Ours  \\
%$$n$ & [smallest, largest] & max & avg & max & avg  & max & avg  & $\frac{\rm Ours}{\cite{modmultqic}}$ & $\frac{\cite{Beckman1996}}{\rm Ours}$ & \#Gates & Depth &  \cite{Beckman1996} & $\frac{\cite{Beckman1996}}{\rm Ours}$  \\
 \hline
 7  & 7 in [65, 119] & 182	& \bf 134.3 &210& \bf 138.3& 1240& \bf 852  & \bf 1.03 & \bf 6.16  &  \bf 1292 & \bf 1083 & 11154 &\bf 8.63\\
 8  & 16 in [133, 253] & 257 & \bf 194.3 &292& \bf 202.8 & 1700& \bf 1162  & \bf 1.04 & \bf 5.73&  \bf 2288 & \bf 2120 & 17415 &\bf 7.61\\
 9  & 34 in [259, 511] & 326 & \bf 258.0 &386& \bf 271.3 & 2232& \bf 1520  & \bf 1.05 & \bf 5.60  &  \bf 3731 & \bf 3059 & 25670 &\bf 6.88\\
 10 & 72 in [515, 1007] & 418 & \bf 327.3 &481& \bf  347.6 & 2836& \bf 1926  & \bf 1.06 & \bf 5.54  &  \bf  5286 & \bf 3885 & 36195 &\bf 6.84 \\
 11 & 152 in [1027, 2047] & 518 & \bf 405.0 &626& \bf 434.5 & 3512& \bf 2380  & \bf 1.07 & \bf 5.48  & \bf 7447 & \bf 4959 & 49266& \bf
  6.61\\
 12 & 299 in [2051, 4087] & 635 & \bf 488.8 &765& \bf 523.8 & 4260& \bf 2882  & \bf 1.07 & \bf 5.50& \bf  10002 & \bf 6075 & 65159 &\bf 6.51 \\
 13 & 621 in [4097, 8189] & 750 & \bf 580.3 &930&  \bf 627.3  & 5080& \bf 3432  & \bf 1.08 & \bf 5.47 & \bf 13364 & \bf 7472 & 84150 & \bf 6.29\\
 14 & 1212 in [8197, 16379] & 882 & \bf 678.6&1120& \bf 738.9 & 5972&  \bf 4030  & \bf 1.08 & \bf 5.45 &  \bf 16854 & \bf 8617 & 106515  & \bf 6.32 \\
 15 &  2429 in [16387, 32765] & - & - & 1340& \bf 868.0 & 6936& \bf 4676  & \bf - & \bf 5.39   &   \bf 21523  & \bf 9985 &132530 & \bf 6.16 \\
 \hline
 \hline
 16 & $M = (2^8-5)\times(2^8-59)$ & - & - &1425 & \bf990.3 & 7972& \bf 5370& \bf - & \bf 5.42     &  \bf 28581 & \bf 15884 & 162471  & \bf 5.68 \\
 \hline
 \hline
 24& $M=(2^{12}-3)\times (2^{12}-77)$ & - & - & 4237 & \bf 2705.9 & 18852 & \bf 12650 &  \bf - & \bf 4.67 & \bf 109405   & \bf 39671 &576455  & \bf 5.27 \\
 32 & $M = (2^{16}-15)\times (2^{16}-123)$ & - & - & 6023 & \bf 5024.0 &34340 & \bf 23002 &\bf - & \bf 4.57 &  \bf 268387 & \bf 86110 &1400679   & \bf 5.21\\
 48& $M=(2^{24}-3)\times (2^{24}-167)$ & - & - & 13447 & \bf 11852.4 & 79140 & \bf 52922 & \bf - & \bf 4.46 &  \bf 954662 & \bf 201065 &4845118   & \bf 5.07 \\
 64 & $M = (2^{32}-5)\times (2^{32}-267)$  & - & - & 24028 & \bf 21354.8 & 142372 & \bf 95130& \bf - & \bf 4.45 &   \bf  2358531&  \bf 422070 &11626238   & \bf 4.92 \\
 \hline
 \hline
 96& $M=(2^{48}-59)\times (2^{48}-257)$ & - & - & \multicolumn{2}{c|}{\bf 46400} & 324132 & \bf 216410 &  \bf -  & \bf 4.66      &  \bf 8.15e6  & \bf 1.00e6 & 3.97e7   & \bf 4.87 \\
 128 & $M = (2^{64}-59)\times (2^{64}-363)$ & -& - & \multicolumn{2}{c|}{\bf 82207} & 579620 & \bf 386842  & \bf -  & \bf 4.71      &    \bf 1.97e7 & \bf 2.03e6 &  9.47e7  & \bf 4.81 \\
 192 & $M = (2^{96}-17)\times (2^{96}-347)$ & -& - & \multicolumn{2}{c|}{\bf 189327} & 1311780 & \bf 875162  & \bf -  & \bf 4.62     &  \bf 6.91e7  & \bf 4.63e6 & 3.22e8 & \bf 4.65 \\
 256 & $M = (2^{128}-159)\times (2^{128}-1193)$ & -& - & \multicolumn{2}{c|}{\bf 331126} & 2338852 & \bf 1560090  & \bf -  & \bf 4.71      & \bf 1.65e8   & \bf 9.25e6 & 7.64e8  & \bf 4.63\\
  384 & $M = (2^{192}-237)\times (2^{192}-1143)$ & -& - & \multicolumn{2}{c|}{\bf 746212} &5277732  & \bf 3519770  & \bf -  & \bf  4.71   &  \bf 5.64e8  & \bf 2.09e7  & 2.59e9   & \bf 4.58\\
 512 & $M = (2^{256}-189)\times (2^{256}-1883)$ & -& - & \multicolumn{2}{c|}{\bf 1324289} & 9396260 & \bf 6265882  & \bf -  & \bf 4.73    & \bf 1.34e9  & \bf 4.12e7 & 6.15e9   & \bf 4.56 \\
\hline
\end{tabular}
\end{table*}

\vspace{-2mm}
\section{Empirical Validation}
\vspace{-2mm}
Our algorithms for on-demand construction of modular multiplication circuits
\footnote{A small fraction of $C$ values are positive or negative modular powers of two, or their modular negations. These rare cases are enumerated directly for each $M$, so that our GCD-based algorithm can skip them.} were embedded into the framework of Figure \ref{fig:shor}.
The number of ancillae in resulting mod-exp circuits was $5n+2$ (as in \cite{modmultqic}), but several optimizations from \cite{modmultqic} were not used, and the number of mod-mult blocks was exactly as in \cite{Beckman1996}. Our software was written in C++ using the GNU MP library (for multi-precision arithmetic) supplied with the GCC 4.6.3 compiler on Linux. We used a workstation with an Intel\textregistered ~Core\texttrademark ~2 Duo 2.2 GHz CPU and 2 GB Memory.

\begin{figure}[b]
\hspace{-5mm}
\includegraphics[width=8cm,height=5.7cm]{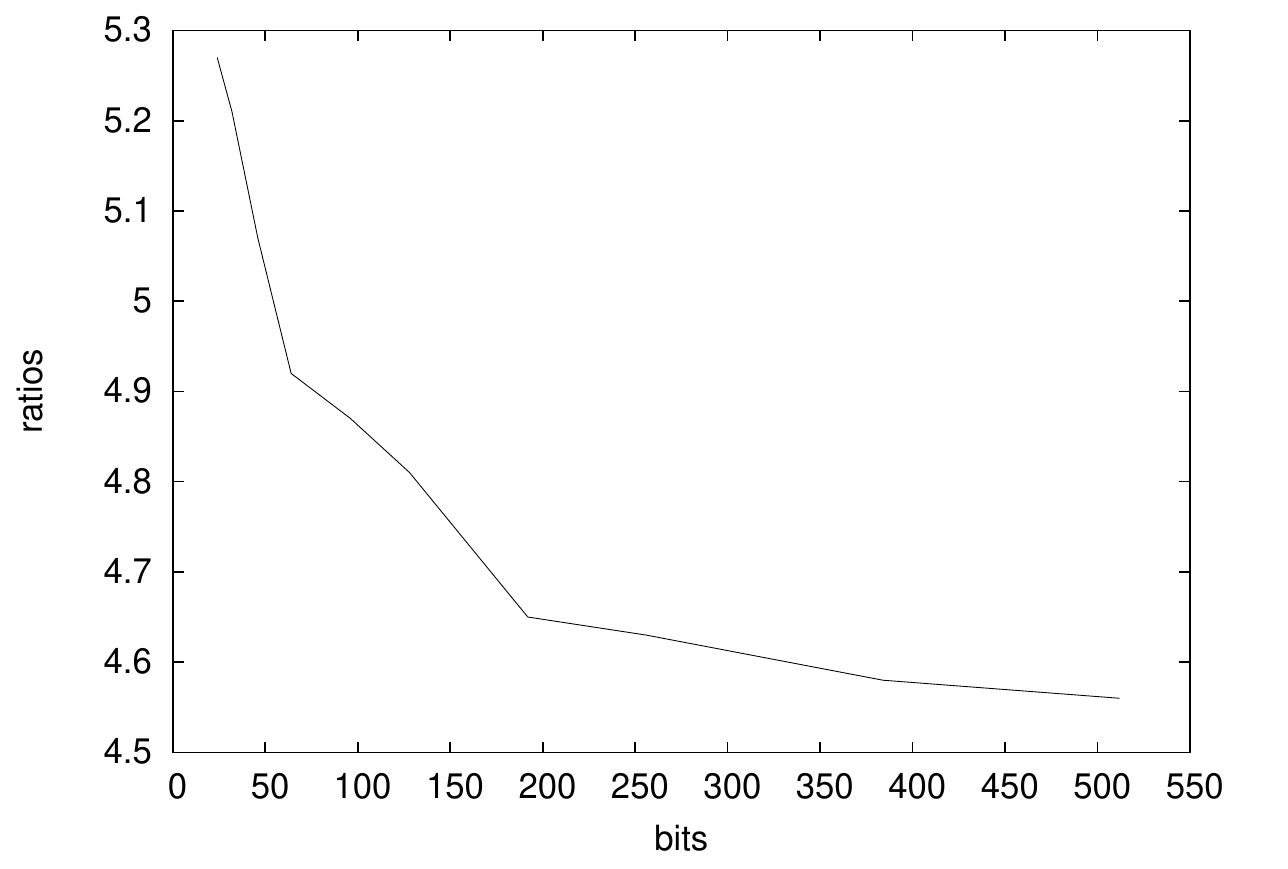}
\vspace{-6mm}
\begin{center}
\caption{\label{fig:asympt}Asymptotic behavior of circuit-size ratios between Beckman et al \cite{Beckman1996} and our constructions.}
\end{center}
\vspace{-8mm}
\end{figure}

To evaluate our optimizations of Shor's number-factoring algorithm, we studied all odd $n$-bit semiprime values of the modulus ($M=pq$) for $7 \leq n \leq 15$, and a subset of $n$-bit $M$ values for $n=$16--512 that are products of the 1st and 10th largest $n$/2-bit primes. Circuit sizes for $n<16$ were averaged over all $M$-coprime $C$ values. Results for $n=16$ include all coprime $C$ values for the given $M$. For 24-, 32-, 48-, and 64-bit $M$ values, results were averaged over the first 5000 coprime $C$ values. For larger $n$ values in modular multiplication circuits, only $C=-1/17 \mod M$ are shown.
Results for modular exponentiation include all $C$ values appearing in unconditional modular multiplication blocks for $b=2$ (Figure \ref{fig:shor}). These are $C=b^{2^0}\%M$, $C=b^{2^1}\%M$, $\cdots$, and $C=b^{2^{2n-1}}\%M$.
For $n\leq 15$, Table \ref{tab:avg_max} shows that circuits found by our heuristic are closer to optimal circuits \cite{modmultqic} than to scalable circuits from \cite{Beckman1996}. Beyond the reach of optimal techniques ($n\geq 24$), Figure \ref{fig:asympt} shows that our circuits are at least 4.5 times smaller and retain their advantage as $n$ increases. Our runtimes ranged from negligible ($n\leq 32$) to 30 min for one 512-bit $(M,C)$ pair.

 To compare our circuits with latency(depth)-optimized constructions in \cite{VanMeter2005}, we note that the most accurate data in \cite{VanMeter2005} are given for $n=128$. Our smallest 128-bit mod-exp circuits use $1.97 \times 10^7$ Toffoli gates with 642 ancillae. To reduce the latency of our circuits, we replaced linear-depth Cuccaro adders with $\log n$-depth adders from \cite{Draper06} also used in \cite{VanMeter2005}. Accordingly, circuit depth is reduced to $2.03 \times 10^6$ Toffoli gates with $\sim900$ ancillae. This process is outlined in the next section, but here we summarize the results. A circuit with 660 ancillae  \cite[Algorithm G, Table II]{VanMeter2005} exhibits latency $1.50 \times 10^7$ Toffoli gates.
 The best circuit in \cite[Algorithm E, Table II]{VanMeter2005} has latency $1.71 \times 10^5$ Toffoli gates but uses 12657 ancillae, which is far less practical with technology under development today. Circuit depths of our modular exponentiation circuits for all attempted $n$ values are reported in Table \ref{tab:avg_max}. A quantum machine with only some limited form of parallelism may still benefit from our techniques, given
 strong results for both parallel and sequential cases.

\vspace{-2mm}
\section{Reducing Circuit Latency}
\vspace{-1mm}
 Our circuits can be adapted to quantum architectures with high degree of parallelism by replacing building blocks by parallelized variants. Circuit-size calculations in Table \ref{tab:avg_max} are based on the costs of circuit modules (addition, subtraction, modular multiplication by 2, etc) from \cite[Table 2]{modmultqic}. Cuccaro adders used in \cite{modmultqic} are small, but exhibit linear latency. To optimize latency for comparisons to \cite{VanMeter2005},
 we replaced Cuccaro adders with QCLA adders from \cite{Draper06} (also used in \cite{VanMeter2005}) whose depth is ($4 \log_2 n + 3$, 4, 2) in terms of (T,C,N) gates.
 As in \cite{VanMeter2005}, we measure latency in Toffoli gates. This results in circuit latency (depth) $4\log_2 n +3$ for additive operators ({\tt \~{}1,\~{}2,+1,+2,-1,-2}) in \cite[Table 2]{modmultqic}. The operators that perform modular multiplication ({\tt d1,d2}) and division by two ({\tt h1,h2}) exhibit latency $6 \log_2 n+ 12$. To count the number of ancillae in our modular exponentiation circuits, note that QCLA adders from \cite{Draper06} need $2n - \log n - 2$ ancillae (vs. 1 for  Cuccaro adders). Given that QCLA adders clear all ancillae, the number of ancillae in our mod-exp circuits grows to $ \le 7n$.

 We also restructure $n$ one-bit controlled-SWAP gates with shared control to reduce latency from $n$ to $\log_2 n$. The control bit is temporarily copied to $n$ zero-initialized ancillae (with $\log_2 n$ latency) \cite{Moore:2002}. We use $n$ parallel one-bit controlled-SWAP gates, and then clear the $n$ ancilla (also with $\log_2 n$ latency). Because these ancillae are cleared immediately, we can share them with the QCLA ancillae. Thus, the overhead is $2 \log_2 n$ latency and $2n$ CNOT gates used to copy/clear ancillae.

\vspace{-2mm}
\section{Conclusions}
\vspace{-2mm}
The $n$-bit multiplication circuits developed in this work significantly simplify the implementation of Shor's algorithm, but use $\Theta(n^2)$ gates, as do traditional circuits.  Circuit sizes are improved by large constant factors. These factors appear exaggerated for {\em small qubit arrays} because, for $b=2$, our construction implements a nontrivial fraction of modular multiplications using $\Theta(n)$ gates using circuit blocks from \cite{modmultqic}. 
In contrast, prior work typically uses generic $\Theta(n^2)$ circuits regardless of $b$.
We have experimented with several enhancements to our technique, but the resulting improvement was not justified by the increased runtime and programming difficulty.

Connections between number factoring and GCD computation were known to Euclid around 300 B.C.E. Today the two problems play similar roles in their respective complexity classes. Number-factoring is in \NP~ (problems whose solutions can be checked in polynomial time), not known to be in \P~(problems solvable in polynomial time), but is not believed to be \NP-complete (most difficult problems in \NP). GCD is in \P, not known to be in \NC~ (problems that can be solved very efficiently when many parallel processors are available), but is not believed to be \P-complete (inherently sequential). Unlike provably-hard problems (such as Boolean satisfiability), or problems for which fast serial and parallel algorithms are known (such as sorting), number factoring and GCD appear to be good candidates for demonstrations of physics-based computing that exploits parallelism.

Our use of GCD algorithms to speed up modular exponentiation and number factoring
incurs only small overhead. 
All the invocations of our GCD-based circuit construction for one run of Shor's algorithm can run in parallel because they are independent. Thus, the classical-computing overhead of our technique for one run of Shor's algorithm is limited to 1-2 GCD-based circuit constructions. This overhead is acceptable because Shor's algorithm performs multiple GCD-like computations after quantum measurement.

\noindent
{\bf Acknowledgments.}
IM's work was sponsored in part by the  Air Force Research Laboratory
under agreement FA8750-11-2-0043. H\'ector J. Garc\'ia helped with Fig. \ref{fig:shor}.

\end{document}

%% file: Shor.tex
\Qcircuit @C=0.5em @R=0.5em {
& \qw & \qw & \qw & \qw & \hdots & & \ctrl{5} & \qw &&&&\\
& \vdots \\
& \qw & \qw & \ctrl{3} & \qw & \hdots & & \qw &  \qw &&&& \\
& \qw & \ctrl{2} & \qw & \qw & \hdots & & \qw & \qw &&&& \\
\\
& \qw {\backslash} & \gate{\left(b^{2^0}\right)\% M } & \gate{\left(b^{2^1}\right)\% M } & \qw & \hdots &  &  \gate{\left(b^{2^{2n-1}}\right)\% M } & \qw &&&&\\
\\
\\
} 

%% file: mux.tex
\Qcircuit @C=0.5em @R=0.5em {
& \qw& \qw & \qw & \qw & \qw & \qw & \qw & \qw & \hdots & & \ctrl{5} & \qw & \ctrl{5} & \qw \\
& \vdots \\
& \qw& \qw & \qw & \qw & \ctrl{3} & \qw  & \ctrl{3} & \qw & \hdots & & \qw & \qw & \qw &  \qw & \\
& \qw& \ctrl{2} & \qw & \ctrl{2} & \qw & \qw  & \qw & \qw & \hdots & & \qw & \qw & \qw & \qw \\
\\
& {\backslash} \qw & \multigate{1}{\rm{SWAP}} & \gate{\left(b^{2^0}\right)\%M } & \multigate{1}{\rm{SWAP}} & \multigate{1}{\rm{SWAP}} & \gate{\left(b^{2^1}\right)\%M }  & \multigate{1}{\rm{SWAP}} & \qw & \hdots &  & \multigate{1}{\rm{SWAP}} & \gate{\left(b^{2^{2n-1}}\right)\%M } & \multigate{1}{\rm{SWAP}} & \qw \\
& {\backslash} \qw & \ghost{\rm{SWAP}} & \qw & \ghost{\rm{SWAP}} & \ghost{\rm{SWAP}} & \qw  & \ghost{\rm{SWAP}} & \qw & \hdots & & \ghost{\rm{SWAP}} & \qw & \ghost{\rm{SWAP}} &  \qw  \\
}

%% file: gcd-mult22.bbl
\begin{thebibliography}{10}

\bibitem{NielsenC2001}
M.~Nielsen and I.~Chuang.
\newblock {\em Quantum Computation and Quantum Information}.
\newblock Cambridge Univ. Press, 2000.

\bibitem{f15}
L.M.K.~Vandersypen et~al.
\newblock Experimental realization of an order-finding algorithm with an {NMR}
  quantum computer.
\newblock {\em Phys. Rev. Lett.}, 85:5452--5455, Dec 2000.

\bibitem{Lanyon2007}
B.~P. Lanyon et~al.
\newblock Experimental demonstration of a compiled version of {Shor's}
  algorithm with quantum entanglement.
\newblock {\em Phys. Rev. Lett.}, 99:250505, Dec 2007.

\bibitem{LuBYP2007}
C.-Y. Lu, D.~E. Browne, T.~Yang, and J.-W. Pan.
\newblock Demonstration of a compiled version of {Shor's} quantum factoring
  algorithm using photonic qubits.
\newblock {\em Phys. Rev. Lett.}, 99:250504, Dec 2007.

\bibitem{PolitiMB09}
A.~Politi, J.~C.~F. Matthews, and J.~L. O'Brien.
\newblock Shor's quantum factoring algorithm on a photonic chip.
\newblock {\em Science}, 325(5945):1221, 2009.

\bibitem{Beckman1996}
D.~Beckman, A.~N. Chari, S.~Devabhaktuni, and J.~Preskill.
\newblock Efficient networks for quantum factoring.
\newblock {\em Phys. Rev. A}, 54:1034--1063, Aug 1996.

\bibitem{modmultqic}
I.~L. Markov, M.~Saeedi.
\newblock Constant-optimized quantum circuits for modular multiplication and
  exponentiation.
\newblock {\em Quantum Info. Comput.}, 12(5-6):361--394, May 2012.

\bibitem{Knill95}
E.~Knill.
\newblock On {Shor's} quantum factor finding algorithm: Increasing the
  probability of success and tradeoffs involving the fourier transform modulus.
\newblock {\em Tech. Report LAUR-95-3350, Los Alamos Natl. Lab}, Aug 1995.

\bibitem{McAnally01}
D.~McAnally.
\newblock A refinement of {Shor's} algorithm.
\newblock {\em arXiv:quant-ph/0112055}, 2002.

\bibitem{Fowler2004}
A.~G. Fowler, S.~J. Devitt, and L.~C.~L. Hollenberg.
\newblock Implementation of {Shor's} algorithm on a linear nearest neighbour
  qubit array.
\newblock {\em Quantum Info. Comput.}, 4(4):237--251, July 2004.

\bibitem{VanMeter2005}
R.~Van~Meter and K.~M. Itoh.
\newblock Fast quantum modular exponentiation.
\newblock {\em Phys. Rev. A}, 71:052320, May 2005.

\bibitem{Takahashi2010}
Y.~Takahashi, S.~Tani, and N.~Kunihiro.
\newblock Quantum addition circuits and unbounded fan-out.
\newblock {\em Quant. Inf. Comput.}, 10(9{\&}10):872--890, 2010.

\bibitem{Rosenbaum12}
D.~Rosenbaum.
\newblock Optimal quantum circuits for nearest-neighbor architecture.
\newblock {\em arXiv:1205.0036}, 2012.

\bibitem{Draper06}
T.~G. Draper, S.~A. Kutin, E.~M. Rains, and K.~M. Svore.
\newblock A logarithmic-depth quantum carry-lookahead adder.
\newblock {\em Quantum Info. Comput.}, 6(4):351--369, July 2006.

\bibitem{Moore:2002}
C.~Moore and M.~Nilsson.
\newblock Parallel quantum computation and quantum codes.
\newblock {\em SIAM J. Comput.}, 31(3):799--815, 2002.

\end{thebibliography}
